# BANSpEmo: A Bangla Emotional Speech Recognition Dataset

Md Gulzar Hussain[1,2,*], Mahmuda Rahman[2], Babe Sultana[2], Ye Shiren[1]

[1]Changzhou University, China
gulzar.ace@gmail.com, yes@cczu.edu.cn
[2]Green University of Bangladesh, Bangladesh
aurthi018@gmail.com, babecse@gmail.com
*Correspondence: gulzar.ace@gmail.com

**Abstract**

*In the field of audio and speech analysis, the ability to identify emotions from acoustic signals is essential. Human-computer interaction (HCI) and behavioural analysis are only a few of the many areas where the capacity to distinguish emotions from speech signals has an extensive range of applications. Here, we are introducing BanSpEmo, a corpus of emotional speech that only consists of audio recordings and has been created specifically for the Bangla language. This corpus contains 792 audio recordings over a duration of more than 1 hour and 23 minutes. 22 native speakers took part in the recording of two sets of sentences that represent the six desired emotions. The data set consists of 12 Bangla sentences which are uttered in 6 emotions as Disgust, Happy, Sad, Surprised, Anger, and Fear. This corpus is not also gender balanced. Ten individuals who either have experience in related field or have acting experience took part in the assessment of this corpus. It has a balanced number of audio recordings in each emotion class. BanSpEmo can be considered as a useful resource to promote emotion and speech recognition research and related applications in the Bangla language. The dataset can be found here: https://data.mendeley.com/datasets/rdwn4bs5ky and might be employed for academic research.*

**Keywords**

*Bangla Speech; Emotion Recognition; Speech Processing; Emotion Classification.*

## VALUE OF THE DATA

- This dataset is the third open-source dataset and can play a significant role in the domain of emotion recognition from audio data for Bangla language.
- It can be used in the emotion recognition from Bangla audio or Bangla language speech recognition or speech-to-text recognition applications.
- Researchers can utilize this dataset by analyzing emotion in Bangla language audio in various fields like psychological examinations, Robotics, mobile services, computer games, contact centers, etc. [1].
- For two sets of sentences, the dataset contains balanced data in each emotion, which might also be used for speech-to-text or vice versa.
- Researchers can use this dataset to train various machine learning models and analyze different features to contribute to Bangla speech emotion recognition (SER).

## 1. DATA DESCRIPTION

Having an audio dataset in any language is very crucial for analyzing emotion from the speech in that language. For languages with low resources like the Bangla language, BANSpEmo is the third audio dataset for emotional speech recognition (SER). BANSpEmo consists of 792 audio recordings of six basic emotional reactions of two sets of sentences. Each set has six sentences. Speakers have explained the emotional states and the utterances have been recorded in a more realistic way than just reading the sentences. These emotional states are Disgust (বিতৃষ্ণা), Happy (খুশি), Sad (দুঃখজনক), Surprised (বিস্মিত), Anger (রাগ), Fear (ভয়). As the emotions of each sentence were explained to the speakers before recording, we can consider the recordings are scripted. The corpus includes voice recordings from 22 students, among them 11 are male and 11 are female. The audio recordings were for two sets of sentences. For recording sentence set 1 (one) 10 males and 8 females were participated and for sentence set 2 (two) only 1 male and 3 females were participated. The gender balance of this dataset has been maintained by including an equal number of male and female speakers, but not for both of the sets of sentences. Therefore, the total number of recordings include (6 sentences × 1 repetition × 6 emotions × 18 speakers) + (6 sentences × 1 repetition × 6 emotions × 4 speakers) = 792 utterances/recordings. The length of the recordings is not fixed, and the length ranges from 3 to 12 sec. This length depends on the sentence size and time taken by the speaker. The total duration of the audio dataset is 1 hour 23 minutes and 12 seconds. Based on the number of audio recordings and the duration of the dataset, this is now the third-largest emotional speech database for Bangla.

## 1.1. Description of Sentences:

There are twelve Bangla sentences splitted in two sets used to prepare BANSpEmo dataset. The six sentences of set one are:

1. কিছু তথ্য সঠিক ভাবে উপস্থাপন করা দরকার, বার বার একই ভুল করে চলেছে সংবাদ মাধ্যম গুলি! (Some information needs to be presented correctly, the media is making the same mistakes over and over again!)
2. আপনার ব্যবহার তো চমৎকার। মুখের ভাষা ও অনেক সুন্দর। (Your behavior is excellent. Your words are also nice.)
3. এর পরিপ্রেক্ষিতে শিক্ষকদের স্বার্থ সংশ্লিষ্ট শিক্ষক সমিতির মধ্য থেকে কোনো ধরনের ভূমিকা পরিলক্ষিত না হওয়ায় আমি ভীষন ভাবে উদ্বিগ্ন । (In this perspective, no role has been observed from the teacher's associations regarding the interest of teachers made me deeply concerned.)
4. আমার একটা ব্যাপার মাথায় ধরে না, "ইলিশ বাঁচাও" স্লোগান মুখরিত মিডিয়া কেন এবং কি কারণে "ইলিশের বাসস্থান (নদী) বাঁচাও" স্লোগান নিয়ে মাতে না? (Why the slogan "Save the habitat (river) of Hilisha" rather than "Save the Hilisha" is being avoided by the media baffles me.)
5. দেশ কি মধ্যম আয়ের দেশে রুপান্তর হচ্ছে নাকি মগের মুলুকের দেশে পরিনত হচ্ছে? (Is the country turning into a middle-income country or a country of anarchy?)
6. আমি একমাত্র সরকারি কোন কাজে আঙ্গুলের চাপ দিতে রাজি আছি, শিক্ষিত ব্যক্তি আঙ্গুলের চাপ দেয় না। (I agree to have my fingerprints used for government purposes, but rational people do not.)

The remaining six sentences of set two are:

1. তগো মনে কতো প্রেম রে জীবনে একটা করছি তাতেই জলে পুরে সেস। (You are bursting with love! I tried once in life but burned myself.)
2. আজ কের ম্যাচ ভারতকে হারাতে চাই টাইগার বাংলাদেশ সাবাস সাকিব আল হাসান। (To defeat India in today's match, we need Bangladesh's tiger, Sakib Al Hasan. )

3. টাইটানিক জাহাজ ডুবে গেছে আর বাংলাদেশ ও ডুবে যাবে । (Bangladesh will go down just like the Titanic did.)
4. প্রশ্ন যদি ভুল হয় তাহলে পরীক্ষা নেবার কি দরকার? সবাইকে গড়ে প্লাস দিয়ে দিবে। ( If the test question is incorrect, why even take it? On average, the highest grade will be given to everyone.)
5. যদি থায় পানতা ইলিশ জুতা দিয়ে তার গালটা কর মালিশ। (He should be punished for doing extravagant expenses during the price hike of Hilisha.)
6. যে জাতি পঁচা ভাত থেয়ে বছর শুরু করে, এরা উন্নতি লাভ করবে কি করে! (No nation will advance by eating soggy rice in the new year!)

The description of the sentences is given in Table 1.

Table 1  Details Description of Used Sentences.

| Specification | Details |
| --- | --- |
| Number of Sets | 2 |
| Number of Sentences in each Set | 6 |
| Total Word Count in Set 1 | 96 |
| Total Word Count in Set 2 | 71 |
| Average words in each Sentence | 14 |

## 1.2. Specification of Audio Files:

BANSpEmo is a class-balanced dataset, and each emotion state has 132 recordings. 16-bit floating WAV format is used to record the audio clips. The Figure 1 illustrates a waveform graph for each emotion of sentence 2 from sentence set 1 and uttered by the speaker 1 from the BANSpEmo dataset.

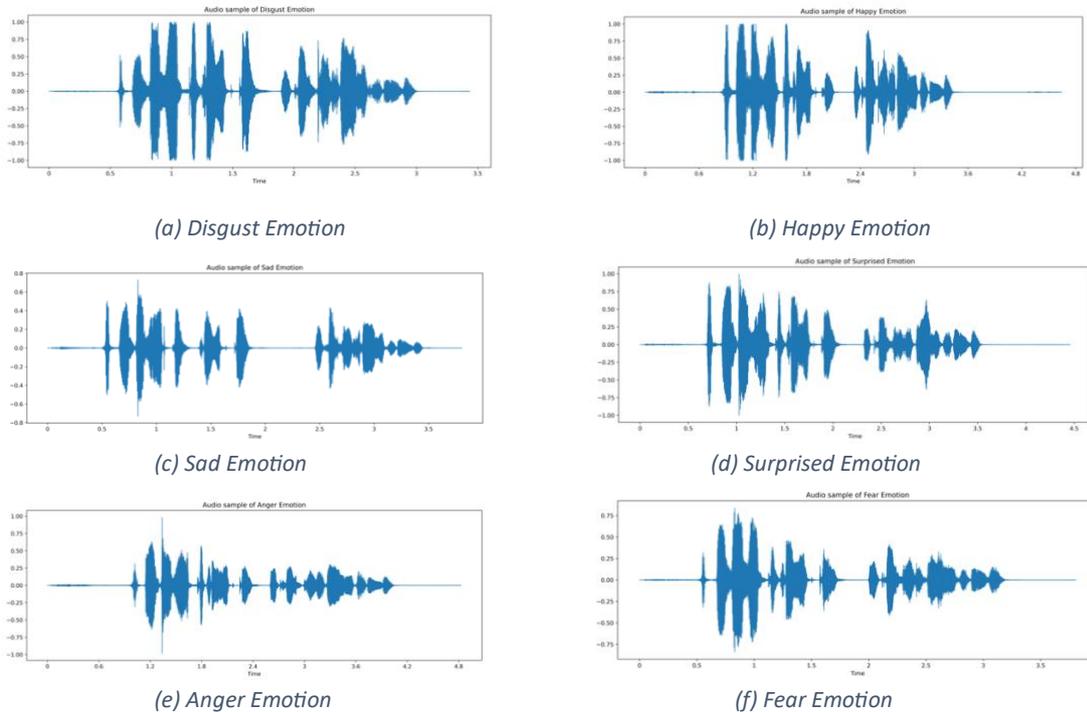

*(a) Disgust Emotion*  *(b) Happy Emotion*
*(c) Sad Emotion*  *(d) Surprised Emotion*
*(e) Anger Emotion*  *(f) Fear Emotion*

Figure 1. Sample Wave-plot of Six Emotions for Sentence 2 from Sentence Set 1 and uttered by the Speaker 1.

In Figure 1, several time periods are shown on the X-axis in seconds, while the signal's magnitude is shown on the Y-axis. Magnitude is the comparative strength of sound waves (transmitted vibrations), which humans experience as loudness. Zero indicates the silence and a value indicates the loudness.

The general range for Y-axis is [-1, 1] but for better visualization, we didn't scale it and used the auto-scaling of the Python Librosa library [2].

## 1.3. SER Datasets in Bangla Language:

There have been a few efforts at developing datasets for speech emotion recognition in the Bangla language. They consist of a small number of recordings and sentences. This restricts the development of a Bangla SER based on deep learning techniques, which frequently require a large number of training inputs. Authors of [3] developed a dataset named ABEG. They used 3 emotional states, which are angry, happy, and neutral. No more description was found about their dataset, and it is not publicly available. A team of researchers created a small private dataset of 160 phrases to test their suggested speech emotion recognition system [4]. 20 speakers who portrayed the emotions of happy, sad, angry, and neutral in this dataset, which lacks perceptual evaluation. However, there are only two publicly available datasets for speech emotion recognition tasks in the Bangla language named SUBESCO [5] and BanglaSER [6]. The most significant feature of BANSpEmo is that, it is a class-balanced, non-preprocessed emotional speech corpus for the Bangla language. A comparison between these public SER datasets in the Bangla language with BANSpEmo is given in Table 2.

Table 2 A comparison Among Various Public Bangla Language SER Corpus and BANSpEmo Dataset.

| Specification | SUBESCO | BanglaSER | **BANSpEmo** |
|---|---|---|---|
| Number of audio records | 7000 | 1467 | 792 |
| Number of Emotions | 7 | 5 | 6 |
| Number of Sentences | 10 | 3 | 12 |
| Number of Performer | 20 | 34 | 22 |
| Professional actors | Yes | No | No |
| Sampling Rate | 48 kHz | 44.1 kHz | 44.1 kHz |
| Class Balanced | Yes | Yes | Yes |
| Gender Balanced | Yes | Yes | No |

## 1.4. SER Datasets in Other Languages:

A review of the literature reveals several noteworthy published studies on the construction and validation of emotional corpora for various languages. A dataset referred to as IEMOCAP [7], was developed by the University of Southern California's Speech Analysis and Interpretation Laboratory. It consists of voice, head movements, and hand movements indicating happiness, anger, frustration, sadness, and neutral condition and has 12 hours length of audio that was verified by six listeners. The audio-visual, multi-modal emotional corpora RAVDESS [8] was recently made available and developed for American English. The 7356 recordings were made by 24 real actors, and more than 250 validators verified their accuracy. EMOVO [9] speech emotion dataset was developed for the Italian language where 6 performers simulated the 6 emotions of anger, disgust, fear, surprise, joy, and sadness. The corpus for the human perception test was evaluated by two separate groups, comprising a total of 24 listeners. Many SER datasets for the languages of Chinese, German, Korean, and Indonesian are available in literature MES-P [10], EMO-DB [11], CADKES [12], IDESC [13], and many other languages. A comparison between some public SER datasets in other languages with BANSpEmo is given in Table 3.

Table 3 Comparison Among Various Public Other Language SER Corpus and BANSpEmo Dataset.

| Specification | RAVDESS | EMOVO | EMO-DB | MES-P | CADKES | **BANSpEmo** |
|---|---|---|---|---|---|---|
| # audio records | 1440 | 588 | 535 | 5376 | 6760 | 792 |
| # Emotions | 8 | 6 | 7 | 7 | 5 | 6 |
| # Sentences | 2 | 14 | 10 | 16 | 52 | 12 |
| # Performer | 24 | 6 | 10 | 16 | 26 | 22 |
| Professional actors | Yes | Yes | Yes | No | No | No |
| Sampling Rate | 48 kHz | 48 kHz | 16 kHz | 44.1 kHz | 16 kHz | 44.1 kHz |
| Class Balanced | No | Yes | No | Yes | Yes | Yes |
| Gender Balanced | Yes | Yes | Yes | Yes | Yes | No |
| Language | English | Italian | German | Chinese Mandarin | Korean | Bangla |

### 1.5. Descriptive Summary of BANSpEmo:

Finally, a descriptive summary of the BANSpEmo dataset is provided in Table 4 for a better understanding of the dataset.

Table 4 Summary of BANSpEmo Dataset.

| Year of Development | 2022 |
|---|---|
| Number of Sets | 2 |
| Language | Bangla |
| Type of Dataset | Scripted |
| Type of Files | Only Audio |
| Audio Clips File Format | .WAV |
| Rate of Sampling | 44.1 kHz |
| Number of Speakers | 22 (11 males and 11 females) |
| Speakers Age Group | 20 to 25 years |
| Number of Emotions | 6 |
| Emotional States | Disgust, Happy, Sad, Surprised, Anger, Fear |
| Number of Sentences | 12 |
| Number of Audio Clips | 792 |
| Size of the Dataset | 915.8 MB |
| Unit Level | Sentence |
| Number of words | 167 |
| The Duration of Each Clip | 3 to 12 s |
| Duration | 1 h 23 min 12 sec |
| Number of Validators | 10 |
| Human Accuracy | 76.05% (approx.) |

### 1.6. Audio Feature Extraction Insights:

Extracting meaningful features from an audio file in order to train any statistical or machine-learning model is very important. Speech feature extraction, which is a crucial stage in the field of audio processing and a branch of signal processing, focuses on this task. It involves the modification or processing of audio data. Transforming digital and analog signals eliminates undesired noise and optimizes the time-frequency spectrum. Features must be carefully chosen in accordance with their impact on model effectiveness. Magnitude Envelope, Zero-Crossing Rate (ZCR), Root Mean Square (RMS) Energy, Spectral Centroid, Band Energy Ratio, and Spectral Bandwidth are a few frequently employed features. Figure 2 shows some aspects of the BanSpEmo dataset's retrieved mel-scaled spectrogram, MFCCs, chroma, Zero-Crossing Rate (ZCR), Tempograph, and Root Mean Square (RMS) features which can be used in future research.

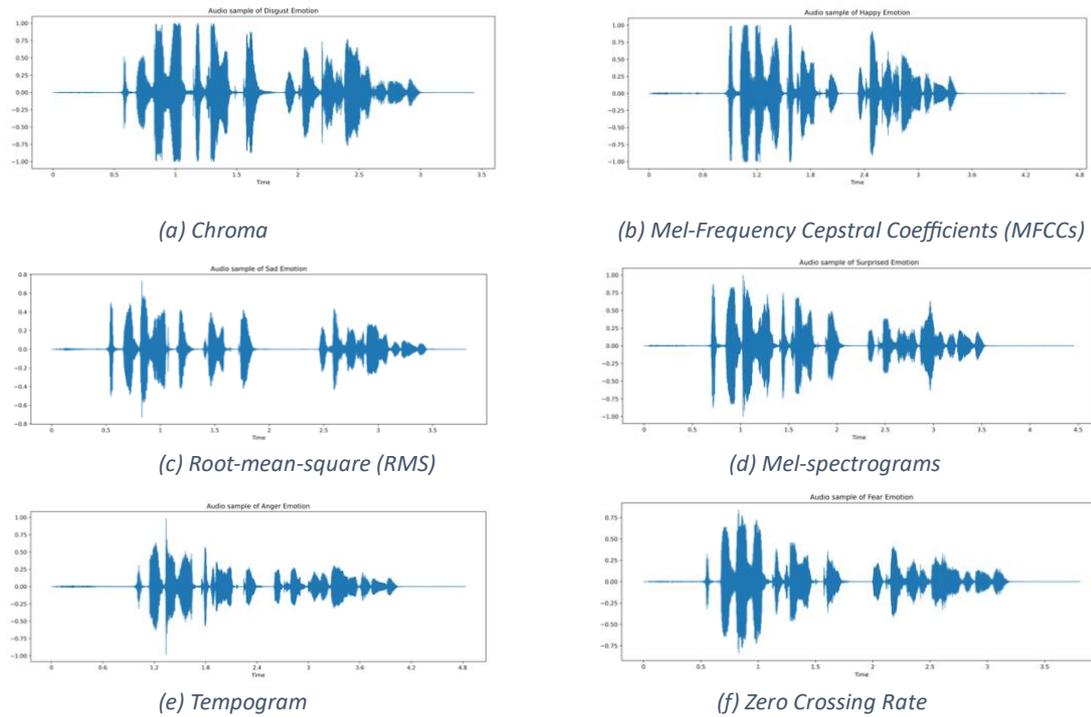

*(a) Chroma*                                *(b) Mel-Frequency Cepstral Coefficients (MFCCs)*

*(c) Root-mean-square (RMS)*                *(d) Mel-spectrograms*

*(e) Tempogram*                             *(f) Zero Crossing Rate*

Figure 2. Sample Features of Happy Emotion for Sentence 2 from Sentence Set 1 and uttered by Speaker 1.

## 2. EXPERIMENTAL DESIGN, MATERIALS AND METHODS

There are five steps in the development of the BANSpEmo dataset: selection of emotions, selection of sentences, briefing the speakers, collection of audio recordings, and evaluation. All of the above steps for assembling the BANSpEmo dataset are shortly explained in this section. Figure 3 displays the workflow diagram for the BANSpEmo development.

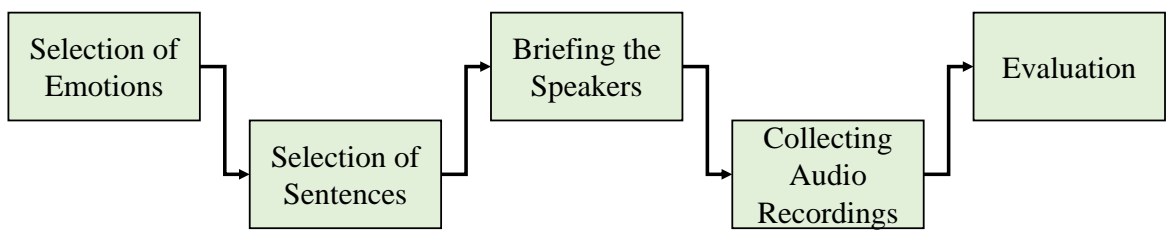

Figure 3. Flowchart for Preparing BANSpEmo Dataset.

### 2.1. Selection of Emotions:

Finding the characteristics to distinguish targeted emotions is the most difficult aspect of emotion recognition. There is still no commonly accepted concept of emotion. Human emotions have always been the subject of numerous divergent theories, both continuous and discrete. Discrete theories involve different quantities and types of fundamental emotions. According to Ekman's approach [14] six fundamental emotions are identified as Disgust (বিতৃষ্ণা), Happy (খুশি), Sad (দুঃখজনক), Surprised (বিস্মিত), Anger (রাগ), and Fear (ভয়). In this research, we selected emotions based on Ekman's theory. This is also remarkably similar to data from other Bangla emotional speech datasets, which can be seen in Table 2. Some hypotheses propose more fundamental feelings like affection, optimism, bravery, or curiosity, but they are considerably harder to express with spoken discourse alone. These fundamental feelings vary in depth and are difficult to define. For instance, because anger and disgust share a high amplitude and pitch issue, the two types of emotions tend to overlap rather frequently. And fear frequently tingles astonishment. So that, the students were given instructions to create emotions that were as autonomous as possible throughout the recording sessions. In each of the six emotions, this dataset contains an equal number of recordings.

## 2.2. Selection of Sentences:

The Bangla text emotion dataset [15] contains simple Bangla language which was collected from various sources like YouTube comments, Facebook comments, blog posts, Bangla textbooks, newspapers, etc. This Bangla text emotion dataset was considered for selecting the text to develop the emotional dataset. For each emotion category, 22 sentences were selected randomly from the emotional text dataset [15]. Then these selected sentences were evaluated by experts who have experience in this domain. After the evaluation of 22 sentences by the experts, 12 sentences are selected and divided into two sets based on the length and emotion level of the sentences. These sentences are as follows:

*Sentence 1:*
  **Bangla**: কিছু তথ্য সঠিক ভাবে উপস্থাপন করা দরকার, বার বার একই ভুল করে চলেছে সংবাদ মাধ্যম গুলি!
  **English Translation**: Some information needs to be presented correctly, the media is making the same mistakes over and over again!
  **Phonemic:** Kichu tathya saṭhika bhābē upasthāpana karā darakāra, bāra bāra ēka'i bhula karē calēchē sambāda mādhyama guli!

*Sentence 2:*
  **Bangla:** আপনার ব্যবহার তো চমৎকার। মুখের ভাষা ও অনেক সুন্দর।
  **English Translation:** Your behaviour is excellent. Your words are also nice.
  **Phonemic:** Āpanāra byabahāra tō camaṭkāra. Mukhēra bhāṣā ō anēka sundara.

*Sentence 3:*
  **Bangla:** এর পরিপ্রেক্ষিতে শিক্ষকদের স্বার্থ সংশ্লিষ্ট শিক্ষক সমিতির মধ্য থেকে কোনো ধরনের ভূমিকা পরিলক্ষিত না হওয়ায় আমি ভীষন ভাবে উদ্বিগ্ন।
  **English Translation:** In this perspective, no role has been observed from the teacher's associations regarding the interest of teachers made me deeply concerned.
  **Phonemic:** Ēra priprēkṣitē śikṣakadēra sbārtha sanśliṣṭa śikṣaka samitira madhya thēkē kōnāē dharanēra bhūmikā parilakṣita nā ha'ōẏāẏa āmi bhīṣana bhābē udbigna.

*Sentence 4:*
  **Bangla:** আমার একটা ব্যাপার মাথায় ধরে না, "ইলিশ বাঁচাও" স্লোগান মুখরিত মিডিয়া কেন এবং কি কারণে "ইলিশের বাসস্থান (নদী) বাঁচাও" স্লোগান নিয়ে মাতে না?
  **English Translation**: Why the slogan "Save the habitat (river) of Hilisha" rather than "Save the Hilisha" is being avoided by the media baffles me.
  **Phonemic:** Āmāra ēkaṭā byāpāra māthāẏa dharē nā, "iliśa bām̐cā'ō" slōgāna mukharita miḍiẏā kēna ēbaṁ ki kāraṇē"iliśēra bāsanthāna (nadī) bām̐cā'ō" slōgāna niẏē mātē nā?

*Sentence 5:*
**Bangla:** দেশ কি মধ্যম আয়ের দেশে রুপান্তর হচ্ছে নাকি মগের মুলুকের দেশে পরিনত হচ্ছে?
**English Translation:** Is the country turning into a middle-income country or a country of anarchy?
**Phonemic:** Dēśa ki madhyama āẏēra dēśē rupāntara hacchē nāki magēra mulukēra dēśē parinata hacchē

*Sentence 6:*
**Bangla:** আমি একমাত্র সরকারি কোন কাজে আঙ্গুলের চাপ দিতে রাজি আছি, শিক্ষিত ব্যক্তি আঙ্গুলের চাপ দেয় না।
**English Translation:** I agree to have my fingerprints used for government purposes, but rational people do not.
**Phonemic:** Āmi ēkamātra sarakāri kōna kājē āṅgulēra cāpa ditē rāji āchi, śikṣita byakti āṅgulēra cāpa dēẏa nā.

*Sentence 7:*
**Bangla:** তগো মনে কতো প্রেম রে! জীবনে একটা করছি তাতেই জ্বলে পুড়ে শেষ।
**English Translation:** You are bursting with love! I tried once in life but burned myself.
**Phonomic:** Tagō manē katō prēma rē! Jībanē ēkaṭā karachi tātē'i jbalē puṛē śēṣa.

*Sentence 8:*
**Bangla:** আজকের ম্যাচ ভারতকে হারাতে চাই টাইগার বাংলাদেশ সাবাস সাকিব আল হাসান।
**English Translation:** To defeat India in today's match, we need Bangladesh's tiger, Sakib Al Hasan.
**Phonemic:** Āja myāca bhāratakē hārātē cā'i ṭā'igāra bānlādēśa sābāsa sākiba āla hāsāna.

*Sentence 9:*
**Bangla:** টাইটানিক জাহাজ ডুবে গেছে আর বাংলাদেশ ও ডুবে যাবে ।
**English Translation:** Bangladesh will go down just like the Titanic did.
**Phonemic:** Ṭā'iṭānika jāhāja ḍubē gēchē āra bānlādēśa ō ḍubē yābē.

*Sentence 10:*
**Bangla:** প্রশ্ন যদি ভুল হয় তাহলে পরীক্ষা নেবার কি দরকার? সবাইকে গড়ে প্লাস দিয়ে দিবে।
**English Translation:** If the test question is incorrect, why even take it? On average, the highest grade will be given to everyone.
**Phonemic:** Praśna yadi bhula haẏa tāhalē parīkṣā nēbāra ki darakāra? Sabā'ikē gaṛē plāsa diẏē dibē.

*Sentence 11:*
**Bangla:** যদি খায় পানতা ইলিশ জুতা দিয়ে তার গালটা কর মালিশ।
**English Translation:** He should be punished for doing extravagant expenses during the price hike of hilisha.
**Phonemic:** Yadi khāẏa pānatā iliśa jutā diẏē tāra gālaṭā kara māliśa.

*Sentence 12:*
**Bangla:** যে জাতি পঁচা ভাত খেয়ে বছর শুরু করে, এরা উন্নতি লাভ করবে কি করে!
**English Translation:** No nation will advance by eating soggy rice in the new year!
**Phonemic:** Yē jāti pam̐cā bhāta khēẏē bachara śuru karē, ērā unnati lābha karabē ki karē!

## 2.3. Speaker Selection and Briefing:

Selecting the appropriate speakers is a vital step in creating an emotional speech database. In order to do this, 36 students from the Green University of Bangladesh who were studying in various disciplines and resided in Dhaka, Bangladesh, were chosen at random to take part in speech-audio recordings. Finally, 22 speakers between the ages of 20 and 25 have been chosen, evenly split between 11 men and 11 women, for the data collection process. These 22 students were selected based on their willingness to participate in the research work voluntarily, and a basic query: if they attended any acting classes? If a student didn't attend any acting classes in his/her past then he/she was selected, because our goal was to record the data from a person who have no acting experience in the past. We selected the students with no acting experience as professional actors might provide an emotion close to the actual state of that emotion. It might create less diversity in the expression of emotions. These students were instructed and trained to deliver their speech as authentically as possible, using their emotions to the fullest extent possible. The sentences and emotional situations were fully explained to the students face-to-face by an expert. The recordings were recorded only after they felt sufficiently prepared.

## 2.4. Collecting of Audio Recordings:

In accordance with the availability of the selected students, we have prepared a schedule and assigned each of them a sequence number. Recording session was held once per day based on their given sequence number. They provided audio files reflecting each of the six emotional moods. Each student has recorded a total of 36 audio recordings, with one repetition for each of the six sentences. To assure quality control, a professional oversaw the entire recording process. The audio recordings were recorded in a supervised laboratory at Green University of Bangladesh's Department of Computer Science and Engineering, where background noise was kept to a minimum. The student was given a seat and given access to a dialog script having numbered sentences. The microphone set-up could be adjusted, which made the recording sessions more convenient and allowed for the best audio capture. Students got enough time to prepare themselves to deliver the right emotions. Each participant received individualized attention as only one student was recorded at a time.

### 2.4.1. Technical information

An appropriate microphone (Boya BY-CM5) mounted on a microphone stand was used by the speaker. The average distance between the microphone and the speaker's mouth was maintained at 7 cm. A USB Audio Interface was attached to the microphone. An Intel(R) Core(TM) i5-10400 CPU running at 2.90GHz with 16.00 GB of RAM was part of the hardware configuration, while Ubuntu 22.04 LTS was the installed operating system. A Dell SE2222H 22-inch Full HD LED VGA HDMI Monitor was additionally supplied to provide visual feedback for audio visualization while the recording was being done. After recording, the recorded voice samples were saved in the ".wav" format with a sampling frequency of 44.1 kHz and 16 bits per sample.

### 2.4.2. Filename Convention

The sentences have been systematically recorded in a predetermined sequential order, with each student receiving a sequence number when they were selected. The student's sequence numbers matched the times they were supposed to be recording. Each data file has received a unique filename according to a predetermined naming structure. Non-numeric identifiers have been used in the filename convention, which is described in Table 5, to indicate the levels of various experimental parameters. The identifiers have been structured as follows: Sentence Set, Sentence Number, Speaker Gender, Speaker Number, Statement Type, Emotion State.wav.

Table 5  Description of Filename Structure of BANSpEmo dataset.

| Identifier | Meaning |
| --- | --- |
| Sentence Set | set 1 - ss1, set 2 – ss2 .... |
| Sentence Number | sentence 1 – s1, sentence 2 – s2 ...... |
| Speaker Gender | male - m, female - f |
| Speaker Number | speaker one – sp1, speaker two – sp2 .... |
| Statement type | Scripted – sc |
| Emotion State | Anger-01, Disgust-02, Fear-03, Happy-04, Sad-05, Surprised-06 |

## 2.5. Corpus Evaluation:

Determining how well untrained listeners understand the emotions being expressed in the recorded audio, is the main goal of corpus evaluation. Better recording quality is indicated by a greater recognition rate. For this dataset preparation, the emotional content was expanded and evaluated by a panel of 10 human validators, made up of 5 men and 5 women. Every validator was either a student or a faculty member from the Green University of Bangladesh. They were selected if they were interested in evaluating voluntarily and had experience in related research fields. Also, they were selected if he/she did not participate in the dataset recording procedure and had acting experience. The speech-audio recording sessions have not been attended by any of the evaluators. All of the validators were fluent readers, writers, and interpreters of Bangla who were native speakers. They did not get any training, especially about the recordings in order to remove any potential bias in their capacity to recognize emotions. Each evaluator gave each recording an evaluation and a corresponding class. Findings are shown in Figure 4. According to the findings, in 76.05% of cases, the validators correctly identified the target emotional class based on their frequently chosen emotions.

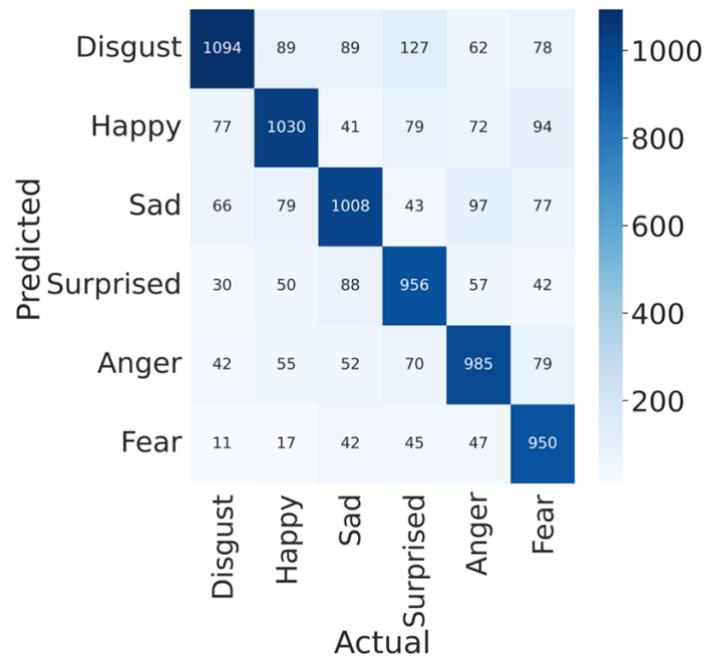

*Figure 4. Corpus Evaluation Confusion Matrix for the Actual vs Predicted Emotions.*

# ETHICS STATEMENT

*Informed consent to release the speech-audio recordings has been acquired from each participating student. Participants were informed that participation was voluntary and that they were free to leave or pause the speech recordings at any time. All data collected from the voluntarily participating students have been anonymized after the full data collection process. It was ensured that any information submitted would be treated confidentially and in an anonymous manner. Each participant has approved the samples post-recording. No ethical approval was required.*

# ACKNOWLEDGEMENTS

*This work was supported in part by the Center for Research, Innovation, and Transformation (CRIT), Green University of Bangladesh [Grant no GUBRG/6/2021].*